\documentstyle[epsf]{mn}

\pagerange{}\volume{}\pubyear{}\date{}

\begin{document}

\title[]{Thermal Relaxation in Young Neutron Stars} 

\author[]{Oleg Y. Gnedin$^1$, Dmitry G. Yakovlev$^{2,3}$, 
          Alexander Y. Potekhin$^2$ \\ 
	  $^1$ Institute of Astronomy, Madingley Road, Cambridge CB3 0HA,
               England; ognedin@ast.cam.ac.uk \\
          $^2$ Ioffe Physical Technical Institute, 
               194021, St Petersburg, Russia \\
          $^3$ Institute for Theoretical Physics,
               University of California, Santa Barbara, CA 93106, USA
}

\maketitle

\begin{abstract}

The internal properties of the neutron star crust can be probed by
observing the epoch of thermal relaxation.  After the supernova
explosion, powerful neutrino emission quickly cools the stellar core,
while the crust stays hot.  The cooling wave then propagates through the
crust, due to its finite thermal conductivity.  When the cooling wave
reaches the surface (age $10-100$ yr), the effective temperature drops
sharply from 250 eV to 30 or 100 eV, depending on the cooling model.
The crust relaxation time is sensitive to the (poorly known) microscopic
properties of matter of subnuclear density, such as the heat capacity,
thermal conductivity, and superfluidity of free neutrons.  We calculate
the cooling models with the new values of the electron thermal
conductivity in the inner crust, based on a realistic treatment of the
shapes of atomic nuclei.  Superfluid effects may shorten the
relaxation time by a factor of 4.  The comparison of theoretical cooling
curves with observations provides a potentially powerful method of
studying the properties of the neutron superfluid and highly unusual
atomic nuclei in the inner crust.

\end{abstract}


\section{Introduction}
  \label{introduc}

Neutron stars are natural astrophysical laboratories of superdense
matter.  In their cores, at densities above the nuclear matter density
$\rho_0 = 2.8 \times 10^{14}$ g cm$^{-3}$, the properties of matter such
as equation of state and even composition are largely unknown (e.g.,
Lattimer \& Prakash 2000, and references therein).  In the absence of
exact theory of superdense matter, different theoretical models predict
different equations of state and compositions (neutrons, protons and
electrons; hyperons; pion or kaon condensates; deconfined quarks).

One of the potentially powerful methods to probe the internal structure
of isolated neutron stars is modelling of their cooling (e.g., Pethick
1992; Page \& Applegate 1992; Page 1998a,b).  The theoretical cooling
curves depend on the adopted physical models of the stellar interior,
especially the neutrino emission and heat capacity, as well as the
superfluidity of neutrons and protons in the core.  Confronting theory
and observations allows one, for example, to constrain the range of
critical temperatures of the superfluidity (e.g., Yakovlev et al.\
1999).

Observing thermal emission of the very young neutron stars, $t \la 100$
yr, opens a possibility of studying the properties of the neutron star
crust.  Soon after a supernova explosion, the young star has large
temperature gradients in the inner parts of the crust.  While the
powerful neutrino emission quickly cools the core, the crust stays hot.
The heat gradually flows inward on a conduction timescale, and the whole
process can be thought of as a cooling wave propagation from the center
towards the surface.  During this thermal relaxation the effective
temperature stays almost constant at about 250 eV.  When the cooling
wave reaches the surface, the effective temperature drops sharply by as
much as an order of magnitude in the fast cooling scenario, and by a
factor of $2-3$ in the slow cooling scenario.  The duration of the
relaxation epoch depends mainly on the heat capacity and thermal
conductivity of the inner crust (Lattimer et al.\ 1994).

Although the equation of state of matter at subnuclear density is known
sufficiently accurately (Negele \& Vautherin 1973, Pethick \& Ravenhall
1995), the properties of atomic nuclei are not.  The nuclei become
unusually neutron-rich, with the smooth proton and neutron
distributions.  At the bottom of the crust, $\rho \ga 10^{14}$ g
cm$^{-3}$, the nuclei can be nonspherical and form clusters (Lorenz et
al.\ 1993, Pethick \& Ravenhall 1995).  The liquid of neutrons dripped
from the nuclei may be superfluid, with the critical temperatures that
are very model-dependent.  The thermodynamic and transport properties of
this matter are subject of large theoretical uncertainty.

In this paper we make improved models of the young neutron stars.  We
obtain the new values of the electron thermal conductivity in the inner
crust, based on a realistic treatment of the shapes of atomic nuclei.
Using a new numerical code, we calculate the cooling models and
determine the duration of the thermal relaxation epoch.  We extend the
analysis of Lattimer et al.\ (1994) and derive the dependence of the
relaxation time on the microscopic parameters of the crust.
Our preliminary results have been summarized by
Yakovlev et al.\ (2000).

\section{Cooling model}
  \label{sect-cool-code}

\subsection{Equations of thermal evolution}

Neutron stars are born very hot in supernova explosions, with the
internal temperature $T \sim 10^{11}$ K, but gradually cool down.  About
twenty seconds after the birth, they become fully transparent for the
neutrinos generated in numerous reactions in stellar interiors.  We 
consider the cooling in the following neutrino-transparent stage.  The
cooling is realized via two channels, by neutrino emission from the
entire stellar body and by heat conduction from the internal layers to
the surface resulting in thermal emission of photons.  For simplicity,
we neglect the possible reheating mechanisms (frictional
dissipation of the rotational energy, Ohmic decay of the internal
magnetic field, or the dissipation associated with weak deviations from
the chemical equilibrium; see, e.g., Page 1998a).

The internal structure of neutron stars can be regarded as
temperature-independent (e.g., Shapiro \& Teukolsky 1983).  The
relativistic equations of thermal evolution include the energy and flux
equations (Thorne 1977):
\begin{equation}
    { 1 \over 4 \pi r^2 {\rm e}^{2 \Phi}} \,
    \sqrt{1 - {2 G m \over c^2 r}} \,
    { \partial  \over \partial  r}
    \left( {\rm e}^{2 \Phi} L_r \right)
    = -Q_\nu - {C_v \over {\rm e}^\Phi} \,
     {\partial T \over \partial t},
  \label{cool-therm-balance}
\end{equation}
\begin{equation}
    {L_r \over 4 \pi r^2} =
    -  \kappa \sqrt{1 - {2Gm \over c^2 r}} \; {\rm e}^{-\Phi}
    {\partial \over \partial r} \left( T {\rm e}^\Phi \right),
  \label{cool-Fourier}
\end{equation}
where $Q_\nu$ is the neutrino emissivity [erg~cm$^{-3}$~s$^{-1}$], $C_v$
is the specific heat capacity [erg~cm$^{-3}$~K$^{-1}$], $\kappa$ is the
thermal conductivity, and $L_r$ is the ``local luminosity" defined as
the non-neutrino heat flux transported through a sphere of radius $r$.
The gravitational mass $m(r)$ and the metric function $\Phi(r)$ are
determined by the stellar model.  After thermal relaxation, the
redshifted temperature $\widetilde{T}(t) \equiv T(r,t)\; {\rm
e}^{\Phi(r)}$ becomes constant throughout the interior.

It is conventional (Gudmundsson et al.\ 1983) to subdivide the
calculation of heat transport in the neutron star interior ($r < R_b$)
and in the outer heat-blanketing envelope ($R_b \leq r \leq R$), where
$R$ is the stellar radius, and the boundary radius $R_b$ corresponds to
the density $\rho_b=10^{10}$ g cm$^{-3}$ ($\sim 100$ meters under the
surface).  The thermal structure of the blanketing envelope is studied
separately in the stationary, plane-parallel approximation to relate the
effective surface temperature $T_s$ to the temperature $T_b$ at the
inner boundary of the envelope.  We use the $T_s$--$T_b$ relation
obtained by Potekhin et al.\ (1997) for the envelope composed mostly of
iron.

The effective temperature determines the photon luminosity: $L_\gamma =
L_r(R,t)=4 \pi \sigma R^2 T^4_s(t)$.  Both $L_\gamma$ and $T_s$ refer to
the locally-flat reference frame on the surface.  A distant observer
would register the ``apparent" luminosity $L_\gamma^\infty = L_\gamma \,
(1 - r_g/R)$ and the ``apparent" effective temperature $T_s^\infty = T_s
\, \sqrt{1 - r_g/R}$, where $r_g = 2GM/c^2$ is the gravitational
radius.  Typically, $r_g/R = 30-40\%$.

We have developed a new evolutionary code, based on the Henyey-type
scheme on a grid of spherical shells (Kippenhahn et al.\ 1967).  The
hydrostatic model of the neutron star with a given equation of state is
calculated separately and is fixed throughout the calculation.  In the
initial configuration the star has a constant redshifted temperature
throughout the interior, $\widetilde{T} = 10^{10}$ K, and no heat flux,
$L_r=0$.  Also, to improve numerical convergence the thermal
conductivity in the core is boosted for the initial epoch $t < 10^{-2}$
yr.  Since the crust is thermally detached from the core at such small
age, this correction has no effect on the cooling curves.  Full details
of the new code are available on the internet
(http://www.ast.cam.ac.uk/$\sim$ognedin/ns/ns.html).

\subsection{Physics input}
  \label{sec:physics}

We use the equation of state of Negele \& Vautherin (1973) in the
stellar crust with the smooth composition model of ground-state matter
to describe the properties of atomic nuclei (Kaminker et al.\ 1999).  We
assume that the nuclei are spherical throughout the entire crust.  The
core--crust interface is placed at $\rho_{cc}=1.5 \times 10^{14}$ g
cm$^{-3}$.  For simplicity, we consider the neutron star cores composed
of neutrons ($n$), protons ($p$) and electrons ($e$) and use the
moderately stiff phenomenological equation of state of Prakash et al.\
(1988) (in the simplified version proposed by Page \& Applegate 1992),
in agreement with our previous work (Yakovlev et al.\ 1999 and
references therein).

\begin{table}
\begin{center}
\caption{Neutron star models}
\begin{tabular}{lllllll}
\hline
      $M$  & $R$  & $\rho_{c14}^c$ 
    & $M_{\rm crust}$ & $\Delta R_{\rm crust}$$^d$ 
    & $M_{\rm D}$ & $R_{\rm D}$  \\
      ($M_\odot$) & (km) &      & ($M_\odot$)
    & (km)  & ($M_\odot$) & (km) \\
\hline
1.1      & 12.20 &  8.50 & 0.050 & 1.66 & \ldots & \ldots \\
1.2      & 12.04 &  9.52 & 0.044 & 1.45 & \ldots & \ldots \\
1.3      & 11.86 & 10.70 & 0.039 & 1.26 & \ldots & \ldots \\
1.4      & 11.65 & 12.20 & 0.033 & 1.09 & \ldots & \ldots \\
1.44$^a$ & 11.54 & 12.98 & 0.031 & 1.02 & 0.000  & 0.00 \\
1.5      & 11.38 & 14.20 & 0.028 & 0.93 & 0.065  & 2.84 \\
1.6      & 11.01 & 17.20 & 0.022 & 0.77 & 0.301  & 4.61 \\
1.7      & 10.37 & 23.50 & 0.016 & 0.59 & 0.685  & 5.79 \\
1.73$^b$ &  9.71 & 32.50 & 0.011 & 0.47 & 0.966  & 6.18 \\
\hline
\end{tabular}
\begin{tabular}{l}
  $^a$  threshold configuration for switching on direct Urca process \\
  $^b$  configuration with maximum allowable mass\\
  $^c$  central density in $10^{14}$ g cm$^{-3}$ \\
  $^d$  $\Delta R_{\rm crust}$ is defined as $R - R_{\rm core}$
\end{tabular}
\label{tab-cool-model}
\end{center}
\end{table}

The parameters of the models are summarized in Table
\ref{tab-cool-model}.  It shows the stellar masses, radii, central
densities, crust masses, and crust thicknesses for a number of models.
The maximum mass of the stable neutron stars with this equation of state
is 1.73 $M_\odot$.  We define the crust thickness as $\Delta R_{\rm
crust}=R - R_{\rm core}$, while the proper geometrical thickness (for
$\Delta R_{\rm crust} \ll R$) is $\Delta R_{\rm crust}/\sqrt{1 -
r_g/R}$.  As we increase the mass $M$, the radii and crust masses of the
stable configurations get smaller, i.e., the stars become more compact.

Free neutrons in the inner crust and both neutrons and protons in the
core of a neutron star are likely to be superfluid.  We assume the
singlet-state pairing of the protons in the core.  Superfluidity of free
neutrons in the crust and of neutrons in the outermost part of the core
is known to be of the singlet-state type, but at higher densities it
switches to the triplet-state type.  
%
%
Various microscopic theories predict a large scatter of the critical
temperatures of the neutron and proton superfluids, $T_{cn}$ and
$T_{cp}$, depending on the nucleon-nucleon potential model and the
many-body theory employed (see Yakovlev et al.\ 1999 for references).

\begin{figure}   
\begin{center}
\leavevmode
\epsfysize=6cm 
\epsfbox[75 270 560 600]{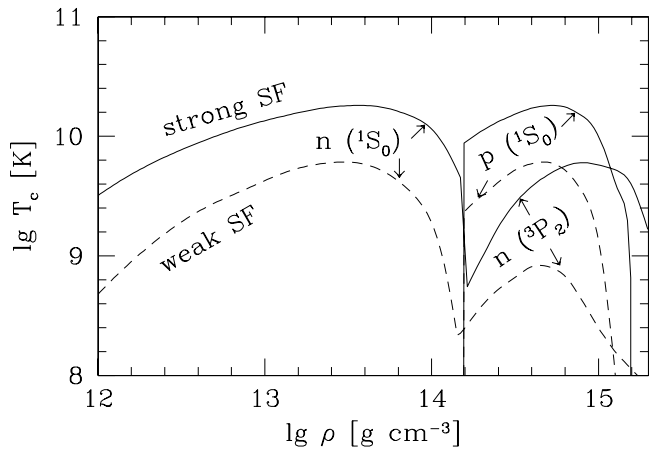}
\end{center}
\caption[ ]{
  Density dependence of the critical temperatures
  of superfluidity (SF) of free neutrons 
  in the inner crust,
  and neutrons and protons in the core
  for the strong (solid lines) and
  weak (dashed lines) superfluid models (see text for details).    
}
\label{fig-cool-tc}
\end{figure}

As an example we will use two models, a {\it weak} and a {\it strong}
superfluidity (Fig.\ \ref{fig-cool-tc}).  The model of strong
superfluidity corresponds to the higher $T_c$.  It is based on the
rather large energy gaps calculated by Elgar{\o}y et al.\ (1996) for the
singlet-state pairing (with the maximum gap of about 2.5 MeV) and by
Hoffberg et al.\ (1970) for the triplet-state pairing.  The weak
superfluid model makes use of the smaller gaps derived by Wambach et
al.\ (1993) (with the maximum gap of about 1 MeV) for the singlet-state
superfluid and by Amundsen \& {\O}stgaard (1985) for the triplet-state
neutron superfluid.  For simplicity, we use the same function $T_c(n)$
to describe the singlet pairing of free neutrons in the crust ($n=n_n$)
and of the protons in the core ($n=n_p$).

\begin{figure}   
\begin{center}
\leavevmode
\epsfysize=8cm 
\epsfbox{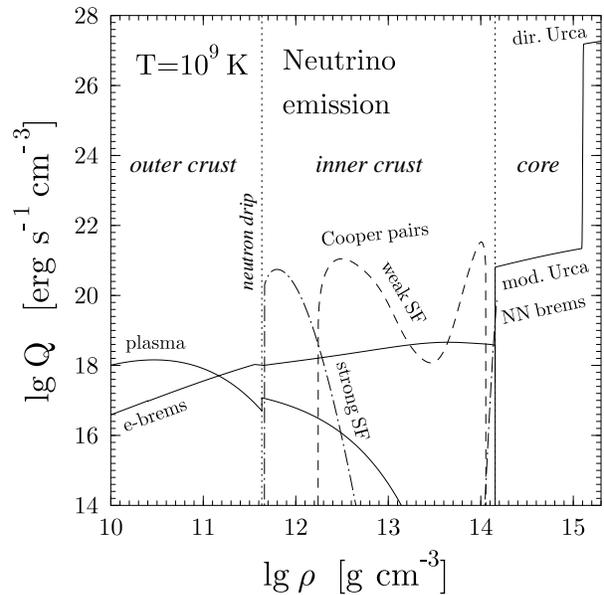}
\end{center}
\caption[ ]{
  Density dependence of the neutrino emissivity
  at $T=10^9$ K.
  Solid lines: partial emissivities due to electron-nucleus
  bremsstrahlung (e-brems) and plasmon decay (plasma) in the crust
  and the total emissivity produced by direct and modified Urca processes
  and by nucleon-nucleon ($N\!N$) bremsstrahlung 
  in the nonsuperfluid core.
  Dashed and dash-and-dot lines: the emissivity due to
  Cooper pairing of neutrons for the models of strong and weak
  superfluidity in the crust.  Vertical dotted lines indicate
  the neutron drip density and the boundary of the core.
}
\label{fig-cool-nu}
\end{figure}

\begin{figure}   
\begin{center}
\leavevmode
\epsfysize=8cm 
\epsfbox{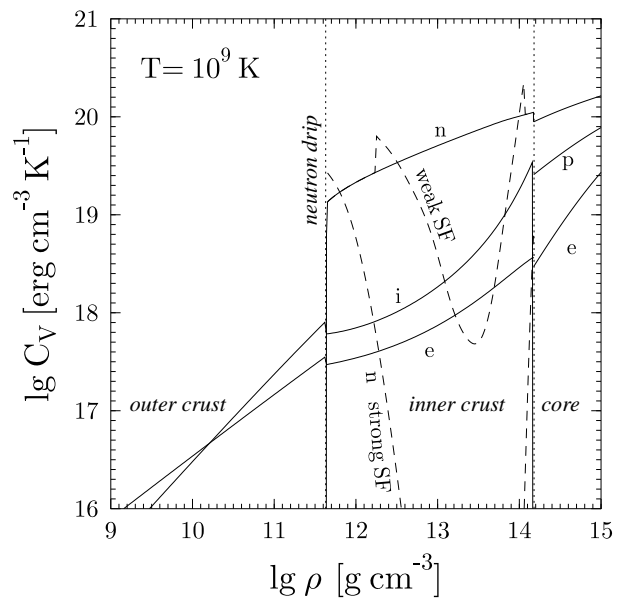} 
\end{center}
\caption[ ]{
  Density dependence of the specific heat capacity
  at $T=10^9$ K.
  Solid lines: partial heat capacities 
  of ions ($i$), electrons ($e$) and free neutrons
  ($n$) in nonsuperfluid crusts, as well as of
  neutrons, protons ($p$) and electrons in nonsuperfluid
  cores. Dashed
  lines: heat capacities of free neutrons
  in the crust modified by weak or strong superfluidity. 
}
\label{fig-cool-c}
\end{figure}

\begin{figure}   
\begin{center}
\leavevmode
\epsfysize=8cm 
\epsfbox{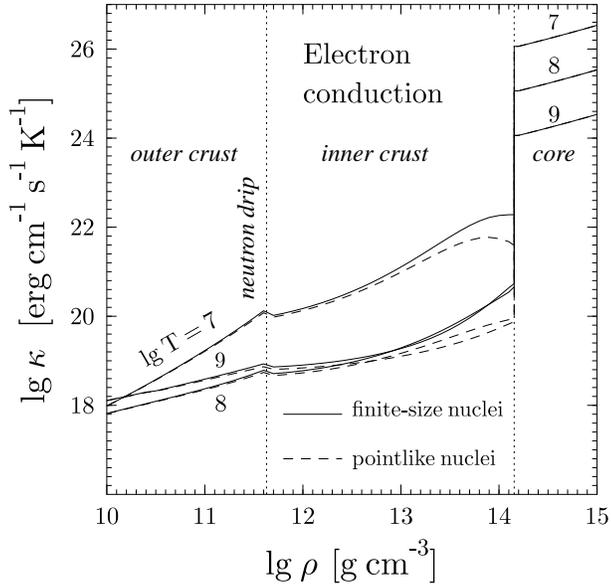} 
\end{center}
\caption[ ]{
  Density dependence of the electron thermal conductivity
  at $T=10^7$, $10^8$ and $10^9$ K in the core and the crust.
}
\label{fig-cool-cond}
\end{figure}

The cooling of neutron stars is mainly determined by the neutrino
emissivity, specific heat capacity, and thermal conductivity.  We
include all relevant sources of neutrino emission (Yakovlev et al.\
1999, 2001): the direct and modified Urca processes, $nn$, $pp$ and $np$
bremsstrahlung in the core; and plasmon decay, $e^-e^+$ pair
annihilation, electron-nucleus ($eZ$) and $nn$ bremsstrahlung in the
crust.  The emissivity of the proton branch of the modified Urca process
derived by Yakovlev \& Levenfish (1995) has been corrected by Yakovlev
et al.\ (2001), which has almost no effect on the cooling curves.  We
include the proper reduction of the neutrino reactions by superfluidity
and also an additional neutrino emission due to Cooper pairing of
superfluid nucleons.  The effective nucleon masses in the core and the
crust are set equal to 0.7 of their bare masses.

A very important cooling effect is produced by the powerful direct Urca
process (Lattimer et al.\ 1991).  For our equation of state, this
process is allowed at densities above the threshold, $\rho_{\rm
crit}=1.298 \times 10^{15}$ g cm$^{-3}$.  If the central density of the
model exceeds the threshold, $\rho_c > \rho_{\rm crit}$, the stellar
core has a central kernel where the direct Urca process leads to fast
cooling.  The masses and radii of these kernels, $M_{\rm D}$ and $R_{\rm
D}$, are given in Table \ref{tab-cool-model}.  The mass of the central
kernel increases rapidly with $M$.  In addition, we show the threshold
configuration with $\rho_c = \rho_{\rm crit}$.  It separates the
low-mass models, where the direct Urca process is forbidden, from the
high-mass models, where the direct Urca is allowed.

For illustration, Fig.\ \ref{fig-cool-nu} shows the emissivity of
various neutrino processes versus density at $T=10^9$ K.  The plasmon
decay is a powerful neutrino emission mechanism in a hot crust at not
very high densities, but it fades away quickly when the temperature
decreases below $10^9$ K.  The electron-nucleus bremsstrahlung is
efficient throughout the entire crust.  In superfluid crusts, the Cooper
pairing neutrino emission switches on at temperatures $T = T_{cn}$,
reaches maximum at $T$ slightly below $T_{cn}$, and fades away
exponentially for $T\ll T_{cn}$.  In the neutron star core, a large jump
of the neutrino emissivity at $\rho=\rho_{\rm crit}$ is associated with
the direct Urca process.
 
The heat capacity is contributed by neutrons, protons and electrons in
the core; and by electrons, free neutrons, and atomic nuclei (vibrations
of ions in Coulomb lattice) in the crust.  The superfluid effects on the
heat capacity of nucleons in the core and of free neutrons in the crust
are incorporated according to Levenfish \& Yakovlev (1994).  In the
absence of superfluidity, the neutrons would have the dominant
contribution in the core and the inner crust.  The effects of neutron
superfluidity are illustrated on Fig. \ref{fig-cool-c} in the case of
crust superfluidity.  When $T$ falls only slightly below $T_{cn}$, the
superfluidity increases the neutron heat capacity due to the latent heat
released at the phase transition.  However, for $T \ll T_{cn}$ the
superfluidity exponentially reduces the heat capacity.  If the stellar
crust was sufficiently colder than in Fig.\ \ref{fig-cool-c}, any
neutron superfluid would reduce the contribution of free neutrons to
negligibly small values, and the heat capacity would be determined by
the ions and electrons.  These effects are analogous in the core.  The
heat capacity in the superfluid core at $T \ll T_{cn}$ and $T \ll
T_{cp}$ would be determined by the electrons.

The thermal conductivity in the core is taken as a sum of the
conductivities of the electrons (Gnedin \& Yakovlev 1995) and neutrons
(Baiko et al.\ 2000).  The electron contribution usually dominates.  The
conductivity in the crust is assumed to be due to the electron
scattering off atomic nuclei.  In the outer crust the finite size of the
proton charge distribution within a nucleus can be neglected.  We use
the recent results of Potekhin et al.\ (1999), which include the
multiphonon processes in electron--nucleus scattering in Coulomb solid
and incipient long-range nucleus--nucleus correlations in strongly
coupled Coulomb liquid of atomic nuclei.  In the inner crust, we have
performed original calculations using the same formalism but taking into
account the finite size of the proton core of atomic nuclei (finite-size
nuclei).  Details of the calculations and the numerical fits are given
in the Appendix.

The results are illustrated in Fig.\ \ref{fig-cool-cond}, which shows
the density profiles of the electron thermal conductivity for $T=10^7$,
$10^8$ and $10^9$ K.  The conductivity in the core is several orders of
magnitude higher than in the crust since there are no such efficient
electron scatterers in the core as atomic nuclei.  In the crust we plot
the thermal conductivity for finite-size nuclei (solid lines) and for
point-like nuclei (dashed lines).  The finite-size effects are
negligible near the neutron drip point but increase the conductivity at
the crust base ($\rho \sim 10^{14}$ g cm$^{-3}$) by a factor of $3-5$.

Another contribution to the thermal conductivity in the crust may come
from the scattering of electrons off charged impurities, the randomly
distributed nuclei of different charge number.  However, the most
important temperature interval for neutron star cooling is $\sim 10^8 -
10^9$ K.  We have verified that the thermal conductivity is almost
unaffected by the impurities in this temperature interval (for a not
very impure matter, $Q_{\rm imp} \la 1$, where $Q_{\rm imp}$ is the
impurity parameter defined in the Appendix), although they can
noticeably decrease the conductivity at lower $T$.  Therefore, we
neglect the effects of impurities in the present calculations.

\section{Thermal relaxation in non-superfluid neutron stars}
  \label{sect-cool-nonsup}

First, consider thermal relaxation in a young non-superfluid neutron
star.  The main features of the process are known from previous cooling
simulations (Page \& Applegate 1992; Lattimer et al.\ 1994; Page 1998a,
1998b; and references therein).

\begin{figure}   
\begin{center}
\leavevmode
\epsfysize=9cm \epsfbox{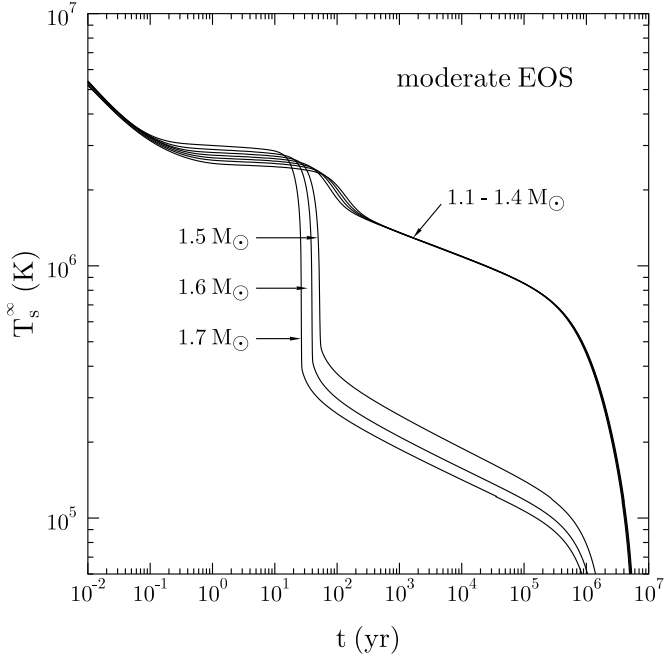}
\end{center}
\caption[ ]{ 
  Cooling curves for non-superfluid neutron star models with 1.1, 1.2
  \ldots, 1.7 $M_\odot$.
}
\label{fig:c_nosf_main}
\end{figure}

\begin{figure}   
\begin{center}
\leavevmode
\epsfysize=9cm \epsfbox{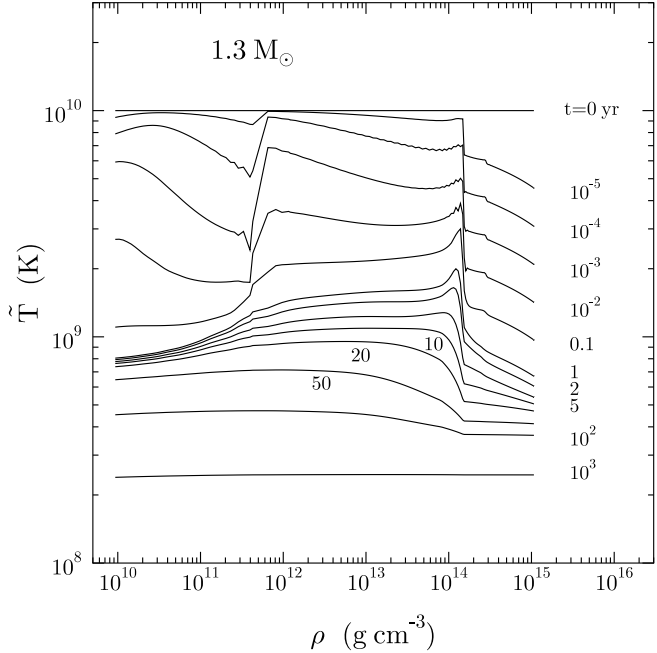}
\end{center}
\caption[ ]{
  Temperature profiles in the 1.3 $M_\odot$ neutron star
  without superfluid effects.
  Numbers next to curves show stellar age.
  Contours are at 0, $10^{-5}$, $10^{-4}$, $10^{-3}$, $10^{-2}$, $10^{-1}$,
  1, 2, 5, 10, 20, 50, 100, and 1000 yr.  
}
\label{fig:config3_nosf}
\end{figure}

\begin{figure}   
\begin{center}
\leavevmode
\epsfysize=9cm \epsfbox{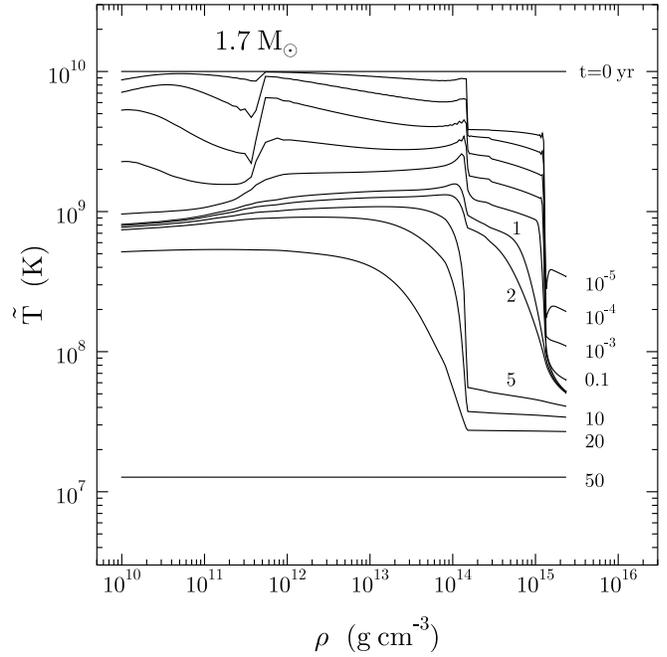}
\end{center}
\caption[ ]{
  Temperature profiles in the 1.7 $M_\odot$ neutron star
  without superfluid effects.
  Contours are at 0, $10^{-5}$, $10^{-4}$, $10^{-3}$, $10^{-2}$, $10^{-1}$,
  1, 2, 5, 10, 20, and 50 yr.  
}
\label{fig:config7_nosf}
\end{figure}

\begin{figure}   
\begin{center}
\leavevmode
\epsfysize=9cm \epsfbox{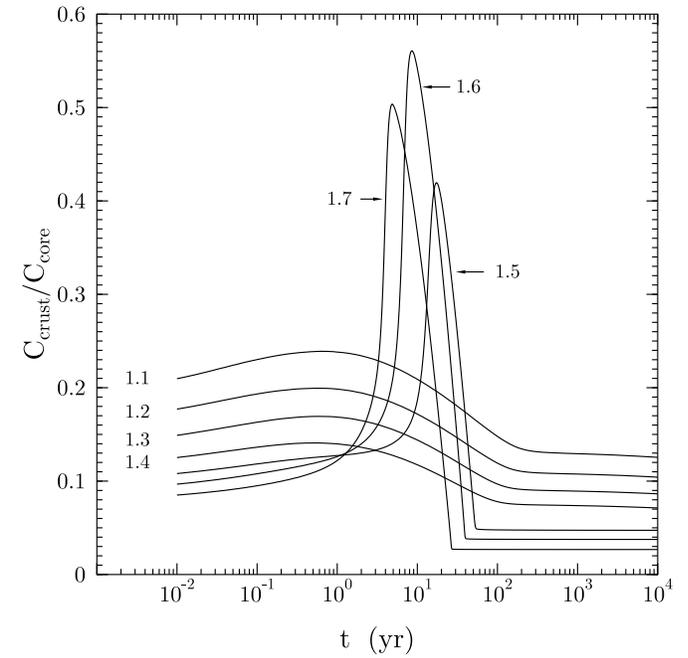}
\end{center}
\caption[ ]{
  Ratio of the integrated heat capacities in the crust and the core for
  the neutron star models of masses 1.1, 1.2, \ldots, 1.7 $M_\odot$.
}
\label{fig:heat}
\end{figure}

\begin{figure}   
\begin{center}
\leavevmode
\epsfysize=9cm \epsfbox{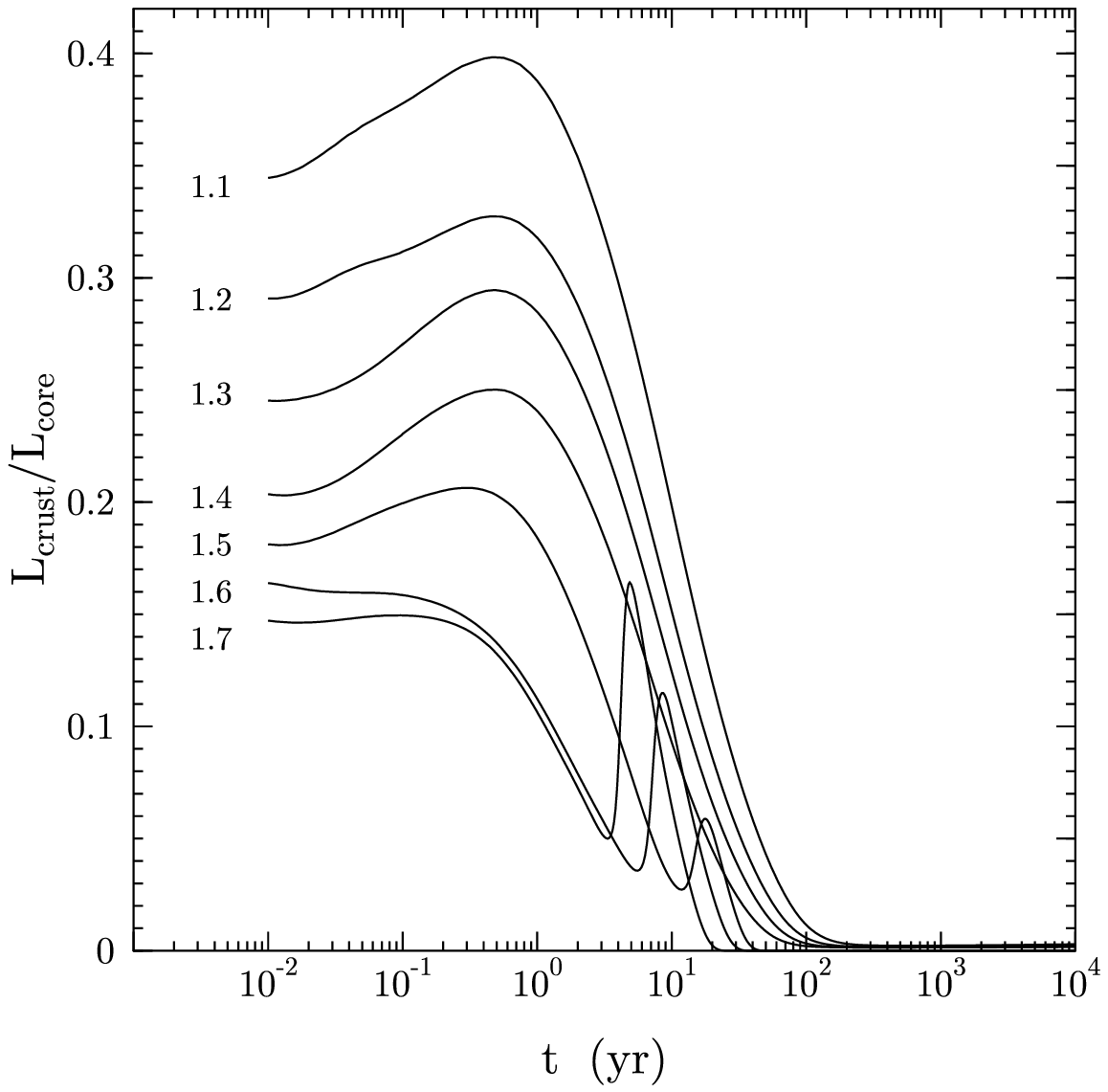}
\end{center}
\caption[ ]{
  Ratio of the neutrino luminosities in the crust and the core for
  the neutron star models of masses 1.1, 1.2, \ldots, 1.7 $M_\odot$.
}
\label{fig:lum}
\end{figure}

Fig. \ref{fig:c_nosf_main} shows cooling models of neutron stars with
different masses.  In the low-mass models, $M < 1.44\ M_\odot$, the
direct Urca process is forbidden.  These stars follow the standard
cooling scenario and the cooling curves are almost independent of $M$.
The high-mass models go through the fast cooling scenario and
demonstrate a spectacular drop of the surface temperature at the end of
the thermal relaxation epoch, $t \sim 50$ yr, due to the emergence of
the cooling wave on the surface.  The same, although much less
pronounced, effect takes place in the case of slow cooling.

Since the neutrino emissivity of the direct Urca process is several
orders of magnitude larger than that of the modified Urca processes, the
fast cooling regime is established even if the central kernel, where the
direct Urca process is allowed, occupies a small fraction of the stellar
core (Page \& Applegate 1992).  For the high-mass stars ($M > 1.44\,
M_\odot$) the cooling curves, again, depend weakly on the mass.  The
change of the slope of the cooling curves at $t \sim 10^5$--$10^6$ yr
manifests the transition from the neutrino to the photon cooling stage.

The surface temperature at the initial cooling stage (the first 100
years) is rather independent of the equation of state, stellar mass, or
the core neutrino luminosity.  The surface temperature is mainly
determined by the physical properties of matter in the crust.  The core
and the crust are thermally decoupled, and the effective surface
temperature does not reflect the thermal state of the stellar core.

In contrast, the evolution of the central temperature, $T(0,t)$, is
drastically different for the slow and fast cooling scenarios.  In all
low-mass models, $T(0,t) \propto t^{-1/6}$ throughout the neutrino
cooling era, $t \la 10^5$ yr, with a small offset in normalization.
This follows from the simple power-law temperature dependence of the
heat capacity ($C_v \propto T$) and the standard neutrino emissivity
($Q_\nu \propto T^8$).  In the models with fast cooling, where the
dominant neutrino process is the direct Urca ($Q_\nu \propto T^6$), the
scaling relation is $T(0,t) \propto t^{-1/4}$ for the initial period $t
< 10^{-2}$ yr.  But then until $t \la 10$ yr, the central temperature
remains almost constant at $10^8$ K as the heat flows from the warmer
outer core, in which direct Urca process is prohibited, into the inner
core.  During the thermal relaxation epoch, $10 < t < 100$ yr, the
central temperature falls again by a factor of several.  After the
thermalization utill the end of the neutrino era, once again, $T(0,t)
\propto t^{-1/4}$.

Figs. \ref{fig:config3_nosf} and \ref{fig:config7_nosf} illustrate the
effects of thermal relaxation on the internal temperature profiles in
the slow and fast cooling scenarios, respectively.  Until the age of
about 1 yr, the neutron star core, the inner and the outer crusts form
almost independent thermal reservoirs.  The region around $4\times
10^{11}$ g cm$^{-3}$, where free neutrons appear in the crust, seems to
be the most effective at cooling owing to the powerful neutrino emission
(see below).  The outer crust cools to $10^9$ K in less than a month,
while the inner parts remain much hotter.  The core also cools
independently but is unable to affect the inner crust due to the slow
thermal conduction.  During the first years the central kernel of the
1.7 $M_\odot$ model in Fig.\ \ref{fig:config7_nosf} remains much colder
than the outer core.  This is because the kernel is cooled by the
powerful direct Urca process and thermal conduction is still unable to
establish thermal relaxation throughout the core.  Almost full core
relaxation is achieved in 10 years.

After the first year, the crust temperature profiles of the slow and
fast cooling scenarios start to differ.  In the former, the temperature
gradient between the core and the crust is slowly eroded, as the cooling
wave from the center reaches the surface.  In the latter, the
temperature gradient continues to grow until it reaches a maximum at $t
\sim 10$ yr.  Then a huge amount of heat releases from the crust and
leads to a spectacular drop of the surface temperature by an order of
magnitude (which lowers the photon luminosity by four orders of
magnitude).  At $t=50$ yr, the entire star is already isothermal.  Note,
that despite larger temperature gradients, thermal relaxation proceeds
overall quicker in the fast cooling scenario.

Prior to thermal relaxation, the contributions of the neutron star crust
to the integrated heat capacity and neutrino luminosity are significant
(Figs.\ \ref{fig:heat} and \ref{fig:lum}).  In the slow-cooling models,
the heat capacity in the crust ranges from 10\% to 20\% of that in the
core, with the larger fraction in the low-mass models (where crusts
occupy larger fraction of the volume).  In the fast-cooling models, the
ratio of the crust to core heat capacities reaches a maximum of 55\% at
$t \sim 10$ yr before dropping to under 10\% after the relaxation.
Similarly, the integrated neutrino luminosity of the crust is about
15\%--40\% of that of the core at $t \sim 1$ yr, and then it drops to a
tiny fraction at later times.

The importance of the individual neutrino mechanisms for the crust
cooling varies at different epochs.  First, for $t < 10^{-2}$ yr in the
fast cooling scenario or for $t < 3\times 10^{-3}$ yr in the slow
cooling scenario, the $e^-e^+$ pair emission controls the crust
temperature.  As the temperature drops below $5\times 10^9$ K, this
process quickly fades away.  The next epoch is controlled by plasmon
decay.  It dominates for $10^{-2} < t \la 10$ yr (fast cooling) or
$3\times 10^{-3} < t \la 10$ yr (slow cooling).  The last epoch of
thermal relaxation lasts for the period $10 \la t < 100$ yr (fast
cooling) or $10 \la t < 1000$ yr (slow cooling), where either
electron-nucleus or neutron-neutron bremsstrahlung is important.  In
fact, both neutrino processes give almost identical cooling curves in
the absence of superfluidity.  However, free neutrons in the crust are
thought to be in a superfluid state which strongly suppresses $nn$
bremsstrahlung.  Therefore, electron-nucleus bremsstrahlung is likely to
be the dominant neutrino mechanism in this last epoch.

\section{Relaxation time}
  \label{sect-reltime}

The duration of the thermal relaxation epoch is potentially interesting
from the observational point of view.  This problem has been studied in
a number of papers, with the most detailed and thorough work by Lattimer
et al.\ (1994; also see references therein).  Those authors considered
thermal relaxation for the fast cooling and defined the relaxation time
$t_w$ as the moment of the most negative slope of the cooling curve,
$\ln{T_s(\ln{t})}$, of a young neutron star.  This is a typical time for
the cooling wave to reach the surface.  We will mainly focus on the case
of rapid cooling, where the relaxation effects are more pronounced.

According to Lattimer et al.\ (1994), the relaxation time of rapidly
cooling neutron stars of various masses is determined mainly by the
crust thickness $\Delta R_{\rm crust}$ and is given by a simple scaling
relation
\begin{equation}
  t_w \approx \alpha \, t_1, \quad \alpha \equiv
  { \left( \Delta R_{\rm crust}  \over 1~{\rm km} \right)^2 }
  \, (1-r_g/R)^{-3/2}.
\label{cool-tw}
\end{equation} 
Here, $t_1$ is the normalized relaxation time which depends solely on
the microscopic properties of matter, such as the thermal conductivity
and heat capacity.  In superfluid neutron stars, $t_1$ is sensitive to
the magnitude and density dependence of the critical temperature of
neutron superfluidity in the crust, as we discuss later.  It is
important that $t_1$ appears to be almost independent of the neutron
star model, its mass $M$ and radius $R$.  We have verified that this
scaling holds also for the slow-cooling non-superfluid models.

For the non-superfluid stars with the core--crust interface placed at
$\rho_0/2$, which is close to our value $\rho_{cc} = 1.5 \times 10^{14}$
g cm$^{-3}$, Lattimer et al.\ (1994) obtained $t_1 \approx 26$ yr.  Our
rapidly cooling models show similar scaling, $t_1 = 28.4 \pm 0.2$ yr,
whereas in the slowly cooling models, $t_1 = 32.9 \pm 1.2$ yr.

The dependence of $t_w$ on the thermal conductivity $\kappa$ and heat
capacity $C_v$ follows from a simple estimate of the thermal relaxation
time in a uniform slab of width $l$:
\begin{equation}
  t_w \sim C_v l^2/\kappa.
\label{eq:slab}
\end{equation} 
The proper width of a thin crust ($\Delta R_{\rm crust} \ll R$), taking
into account the effects of general relativity, is $l= \Delta R_{\rm
crust}/\sqrt{1-r_g/R}$.  This gives $t_w \propto 1/(1-r_g/R)$ in
equation (\ref{cool-tw}).  An additional factor $1/\sqrt{1-r_g/R}$
accounts for the gravitational dilation of time intervals.

The parameter $t_1$ can be roughly estimated using equation
(\ref{eq:slab}) with the values of the specific heat capacity and
thermal conductivity in the crust near the crust--core interface, for
instance, at $\rho = 10^{14}$ g cm$^{-3}$.  Fig. \ref{fig:config7_nosf}
shows that the relevant temperature at the interface is $T \approx
2\times 10^8$ K.  At those temperature and density according to our
physics input, $C_v = 1.9 \times 10^{19}$ erg cm$^{-3}$ K$^{-1}$ and
$\kappa = 1.8 \times 10^{20}$ erg cm$^{-1}$ s$^{-1}$ K$^{-1}$.  This
gives $t_1 \sim (1~{\rm km})^2 \, C_v/\kappa \approx 34$ yr, in
qualitative agreement with the model value.

\begin{table}
\caption{Relaxation time $t_w$  and normalized time $t_1$ for neutron stars
         with different crust models}
\begin{center}
\begin{tabular}{lrrr}
\hline
test crust model                & $t_w^a$ (yr) & $t_1^a$ (yr) & $t_1^b$ (yr) \\
\hline
real model, no SF                         &   52.4  &   28.8  &   33.9  \\
\quad no crust neutrinos                  &  253.5  &  139.2  &  134.9  \\
\quad only plasmon decay neutrinos        &   67.6  &   37.1  &   41.6  \\
\quad only $eZ$ neutrino bremsstr.        &   58.3  &   32.0  &   34.5  \\
\quad no neutron heat capacity            &   15.3  &    8.4  &    6.7  \\
\quad cond. for point-like nuclei         &  131.8  &   72.4  &  102.3  \\
real model, weak crust SF                 &   20.2  &   11.1  &    3.3  \\
\quad no Cooper neutrinos                 &   29.0  &   15.9  &   19.0  \\
\quad weak core+crust SF                  &   22.3  &   12.2  &   25.7  \\   
real model, strong crust SF               &   15.0  &    8.2  &    6.7  \\
\quad no Cooper neutrinos                 &   15.5  &    8.5  &    6.9  \\
\quad strong core+crust SF                &   10.7  &    5.9  &    5.8  \\
\hline
\end{tabular}
\label{tab:lab}
\begin{tabular}{l}
  $^a$  for the 1.5 $M_\odot$ model, with $\alpha=1.821$ \\
  $^b$  for the 1.3 $M_\odot$ model, with $\alpha=2.875$
\end{tabular}
\end{center}
\end{table}

\begin{figure}   
\begin{center}
\leavevmode
\epsfysize=9cm \epsfbox{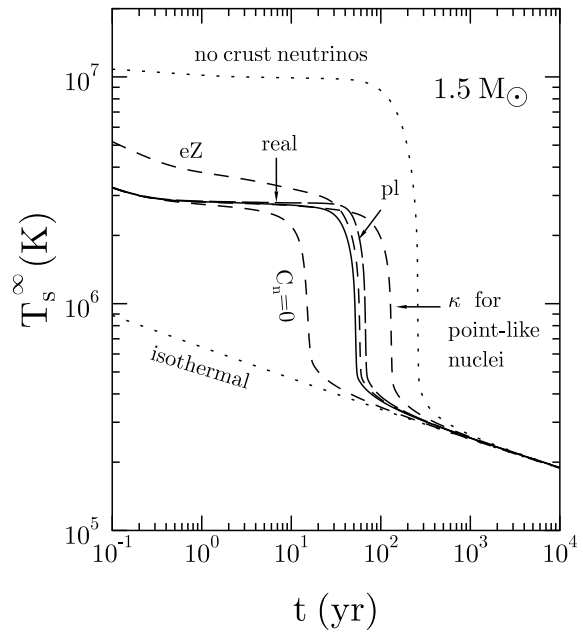}
\end{center}
\caption[ ]{
  Thermal relaxation for the 1.5 $M_\odot$ model without
  superfluid effects (see also Table \protect\ref{tab:lab}).  
  Solid line: the real cooling
  curve. Dotted lines: switched off neutrino emission
  from the crust (upper) or infinite thermal
  conductivity at $\rho > 10^{10}$
  g cm$^{-3}$ (lower). The dashed curve $C_n=0$: 
  removed neutron heat capacity in the crust.  Another
  dashed curve: the thermal
  conductivity $\kappa$ in the crust is for point-like nuclei.
  Two other dashed lines:
  removed all neutrino
  mechanisms in the crust except either
  plasmon decay (pl) or electron-nucleus
  bremsstrahlung ($eZ$).
}
\label{fig:c5_cr}
\end{figure}

Figure \ref{fig:c5_cr} shows the sensitivity of the relaxation time to
test variations of physical properties of the neutron star crust in the
fast cooling scenario of the 1.5 $M_\odot$ star.  The corresponding
values of $t_w$ and $t_1$ are listed in Table \ref{tab:lab}.  They may
differ from the mean values (Table \ref{tab:fits}) within the
error-bars.  We present also the results for the superfluid models
discussed in the next section.

Switching off the neutrino emission from the crust, while keeping the
heat capacity, slows down the thermalization epoch by almost an order of
magnitude, from 50 to 250 yr.  Turning on the plasmon decay alone would
lead to the cooling curve not very different from the real one, although
the thermalization epoch would be delayed by about 30\% ($t_w = 68$ yr).
Turning off the plasmon decay and switching on the electron-nucleus
bremsstrahlung instead would give a hotter neutron star before the
relaxation but almost correct relaxation time, $t_w = 58$ yr.  Combined
together, plasmon decay and electron-nucleus bremsstrahlung would
reproduce accurately the real cooling curve.

Restoring the neutrino emissivity in the crust but switching off the
heat capacity of crustal neutrons ($C_n=0$) we obtain much faster
relaxation, which lasts about 15 years.  This numerical experiment
imitates the suppression of the neutron heat capacity by strong neutron
superfluidity (discussed in Sect.\ \ref{sect-cool-super}).  On the other
hand, had we neglected the quantum suppression of the heat capacity of
the nuclei below the Debye temperature, the latter heat capacity would
have become important in older neutron stars, $t \ga 10^4$ yr, strongly
delaying the cooling.

Finally, the relaxation time depends on the thermal conductivity in the
inner crust.  For instance, a neglect of the finite sizes of atomic
nuclei in the electron--nucleus scattering rate would lower the electron
thermal conductivity at the crust base ($\rho \ga 10^{13}$ g cm$^{-3}$)
by a factor of 2--5 (cf Fig.\ \ref{fig-cool-cond}).  Using that, less
realistic thermal conductivity we would have had much longer relaxation
(about 130 years).  If the thermal conductivity were infinite in the
stellar interior (at $\rho > \rho_b$), we would have obtained an
isothermal cooling scenario.  In this case a sharp drop of the surface
temperature associated with the relaxation disappears.

We have run additional cooling models with $M = 1.5 \,
M_\odot$, varying the heat capacity and thermal conductivity within the
crust (at $\rho_b \leq \rho \leq \rho_{cc}$) by a fixed factor 1/8, 1/4,
1/2, 2, 4, 8.  The results confirm that $t_w$ and $t_1$ are indeed quite
sensitive to the variations of $C_v$ and $\kappa$, in agreement with the
qualitative estimate, equation (\ref{eq:slab}), and the results of
Lattimer et al.\ (1994).  It is important that these variations do not
invalidate the scaling relation for the relaxation time, equation
(\ref{cool-tw}).  Moreover, if $C_v$ is increased and $\kappa$ is
decreased, the dependence of $t_1$ on the values of the heat capacity
and thermal conductivity is described by a simple scaling relation $t_1
\propto C_v/\kappa^{0.8}$.

\begin{figure}   
\begin{center}
\leavevmode
\epsfysize=9cm \epsfbox{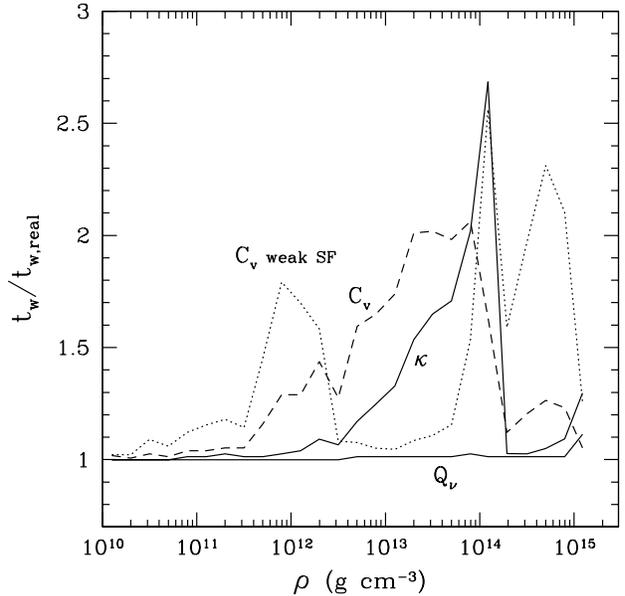}
\end{center}
\caption[ ]{
  Sensitivity of the crust relaxation time to the variations of
  the heat capacity, thermal conductivity, and neutrino emissivity
  in various density regions of the 1.5 $M_\odot$ neutron star
  model with the non-superfluid core.  In each 0.2 dex of $\log{\rho}$,
  $Q_\nu$ and $\kappa$ are reduced by a factor of 8 (solid lines,
  no superfluidity), and $C_v$ is enhanced by a factor of 8 without
  superfluid effects (dashes) and with weak superfluidity (dots).
}
\label{fig:twvar}
\end{figure}

We have also done sensitivity tests of the relaxation time, analogous to
those performed by Epstein et al.\ (1983).  In each density region, 0.2
dex of $\log{\rho}$, either the heat capacity, thermal conductivity, or
neutrino emissivity have been changed by a factor 1/8, 1/4, 1/2, 2, 4,
8.  Figure \ref{fig:twvar} shows the variations of $t_w$ when $Q_\nu$
and $\kappa$ are reduced by a factor of 8, and $C_v$ is enhanced by a
factor of 8.  The relaxation time depends mostly on the values of $C_v$
and $\kappa$ in the crust in the density range $10^{13} - 1.5\times
10^{14}$ g cm$^{-3}$ near the crust-core interface, being rather
insensitive to the variations of $Q_\nu$.  Variations of the physical
parameters in the core affect the crustal relaxation much less strongly
(at least for non-superfluid models).  The density range where $t_w$ is
most affected by the variations of $\kappa$ is narrower than the density
range where it is affected by the variations of $C_v$.  The most
important temperature range which influences $t_w$ is $10^8 \la T \la
10^9$ K.  The properties of matter in these ``sensitivity strips'' of
$\rho$ and $T$ are very model-dependent.  For instance, the nuclei may
be strongly non-spherical (rods, plates, etc.) at $\rho > 10^{14}$ g
cm$^{-3}$, which is not included in our calculations.  Note that the
thermal conductivity has not been calculated so far for the phase of
non-spherical nuclei.

However, if $C_v$ were noticeably lower or $\kappa$ noticeably higher
than in our basic non-superfluid models, the decrease of the crust
relaxation time would saturate at $t_w \approx 13$ yr.  This is the time
it takes the inner core with the direct Urca emission to equilibrate
thermally with the outer core (cf Fig.\ \ref{fig:config7_nosf}).  More
generally, this is a core relaxation time $t_{\rm core}$, which is
almost independent of the parameters of the crust.  It can be estimated
using the same formalism of heat diffusion, equation (\ref{eq:slab}),
through a slab of material between the direct-Urca-allowed kernel and
the boundary of the core, with $l = R_{\rm core}-R_D$.  The relativistic
factors appear as $(e^{-\Phi_{cc}})^3$, where $\Phi_{cc}$ is the metric
function at the core-crust interface.

We have calculated the core relaxation time by setting the crust
conductivity very high for the fast cooling models of non-superfluid
neutron stars with $M=1.5, 1.6, 1.7\, M_{\sun}$.  The least-squares fit
gives
\begin{equation}
  t_{\rm core} = t_2 \;
                 \left({ R_{\rm core}-R_D \over 10\ \mbox{km} }\right)^2
                 e^{-3 \Phi_{cc}},
  \label{eq:tcore}
\end{equation} 
with $t_2 =(9.1 \pm 0.8)$ yr.  Thus, the fast cooling models may have
two distinct relaxation times, in the core and in the crust, and the
latter is typically longer, at least for non-superfluid models.

The normalized crustal relaxation time $t_1$ shows very small variations
with the neutron star mass, $\pm 0.2$ yr for fast cooling and $\pm 1.2$
yr for slow cooling (see Table \ref{tab:fits}).  Even combining the
models with all masses we find that the scaling relation,
eq. (\ref{cool-tw}), holds remarkably well, $t_1 (\mbox{all}) = 31 \pm
2$ yr.  Therefore, the cooling of the crust prior to thermal relaxation
is indeed insensitive to the thermal evolution of the core.  The slow
cooling makes the relaxation much less pronounced but it does not change
its basic features.  This property owes to the very high thermal
conductivity in the core (Fig.\ \ref{fig-cool-cond}), which makes the
core relaxation time typically much shorter than the crust one.
Accordingly, the cooling curves do not depend on the exact values of the
conductivity in the core as long as it is much higher than in the crust.

\begin{figure}   
\begin{center}
\leavevmode
\epsfysize=9cm \epsfbox{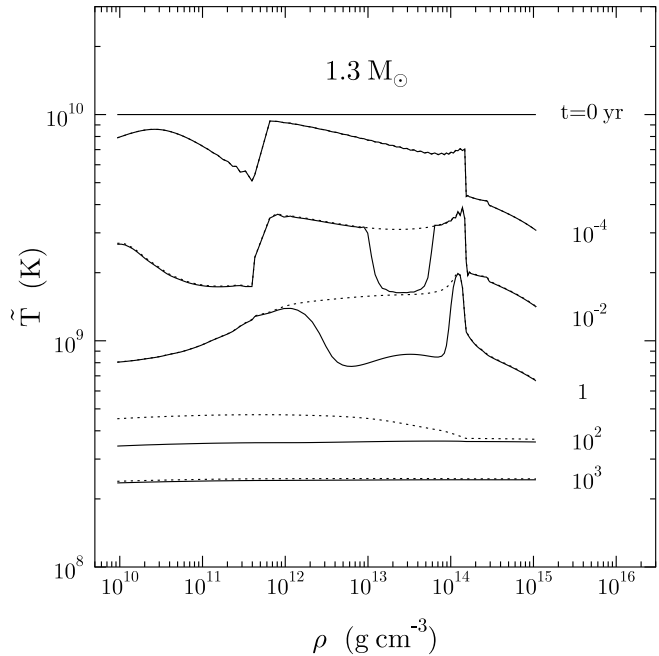}
\end{center}
\caption[ ]{
  Temperature profiles in the interior of the 1.3 $M_\odot$ model
  with (solid lines) and without (dots) weak crust superfluidity
  of free neutrons.  The core is not superfluid.
  Numbers next to curves show the stellar age.
  The contours are at 0, $10^{-4}$, $10^{-2}$, 1, 100, and 1000 yr.
}
\label{fig:config3_crsf2}
\end{figure}

\begin{figure}   
\begin{center}
\leavevmode
\epsfysize=9cm \epsfbox{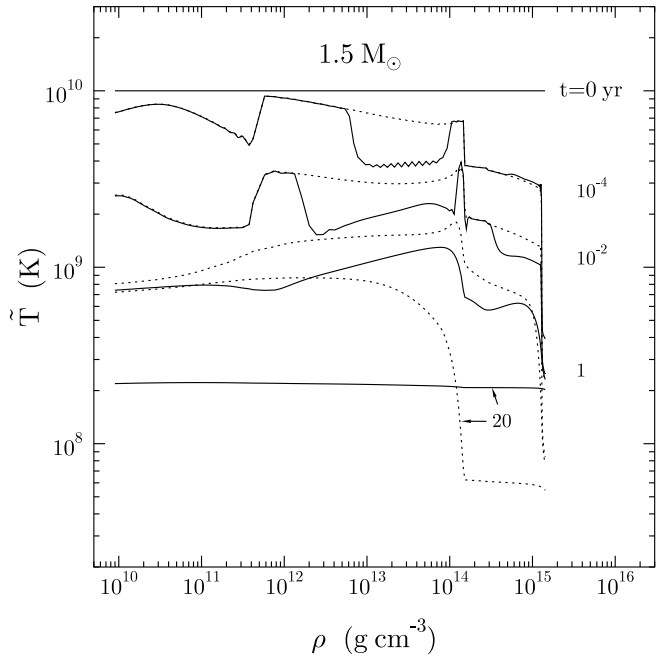}
\end{center}
\caption[ ]{
  Temperature profiles (solid lines) in the interior of the 1.5 $M_\odot$
  model with strong superfluidity both in the crust and the core.
  Numbers next to curves show the stellar age.  The contours are at
  0, $10^{-4}$, $10^{-2}$, 1, and 20 yr.  
  Dotted lines show the temperature profiles of the non-superfluid star.
}
\label{fig:config5_sf1}
\end{figure}

\begin{figure}   
\begin{center}
\leavevmode
\epsfysize=8.5cm \epsfbox{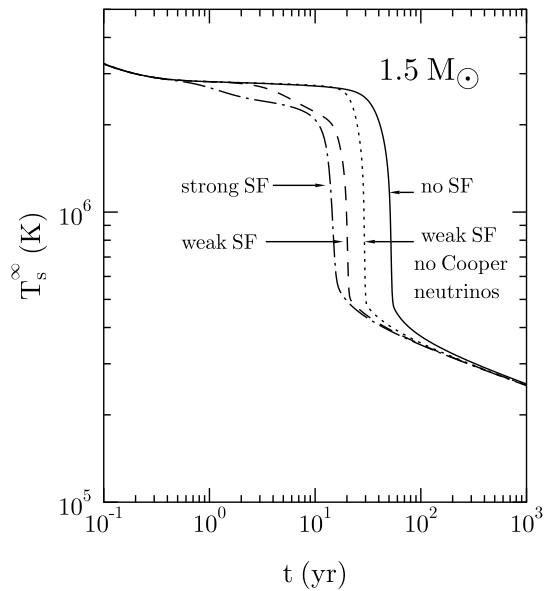}
\end{center}
\caption[ ]{
  Superfluid effects in the crust of the 1.5 $M_\odot$ neutron star
  model with the non-superfluid core.  Dashed line is for the case of
  weak neutron superfluidity, while dash-and-dot line is for the
  case of strong superfluidity.  Dotted line is obtained
  neglecting the Cooper-pair neutrino emission for the weak-superfluid model.
  Solid line is the cooling curve for the non-superfluid crust.
}
\label{fig:c5_crsf}
\end{figure}

\section{Effects of superfluidity}
  \label{sect-cool-super}

Free neutrons in the crust and both neutrons and protons in the core of
a neutron star are likely to be in the superfluid state.  We implement
the effects of superfluidity as discussed in Sect.\ \ref{sec:physics}.
It turns out that superfluidity in the crust affects the cooling curves
at the thermal relaxation stage, while superfluidity in the core affects
cooling at later stages.  First, consider the effects of neutron
superfluidity in the crust.  If the temperature $T$ falls much below the
critical temperature $T_{cn}$, the superfluidity reduces strongly the
neutron heat capacity and $nn$ neutrino bremsstrahlung.  While the
latter is compensated by electron-nucleus bremsstrahlung, the former
effect leads to a faster thermal relaxation.  In addition, a new
neutrino process is allowed in the superfluid state, the neutrino
emission due to Cooper pairing of free neutrons.  It further accelerates
cooling and thermal relaxation of the crust.

Figure \ref{fig:config3_crsf2} illustrates how the weak neutron
superfluidity in the crust carves out the temperature profiles in the
standard cooling scenario, $M = 1.3 \, M_\odot$.  For the first
$10^{-4}$ yr, when the temperature is above $T_{cn}$, the thermal
structure is identical to that of the non-superfluid model.  Later, the
region where the critical temperatures are the highest cools much faster
than the neighboring layers.  The acceleration of cooling is mainly due
to the reduction of the heat capacity and switching on the
Cooper-pairing neutrino emission.  A sequence of points in Fig.\
\ref{fig:config3_crsf2} in which solid lines start to deviate from the
dotted ones reproduces the density profile of $T_{cn}$ shown in Fig.\
\ref{fig-cool-tc}.  As the temperature falls further, wider density
regions become affected, producing shells of cool matter surrounded by
hotter layers on both sides.  After the cooling wave from the core
reaches the outer crust, the star settles into almost the same
isothermal state as the non-superfluid model, but faster.

The effects are much stronger if the superfluidity is allowed for in the
stellar core.  Figure \ref{fig:config5_sf1} shows the combined effect of
the strong superfluidity in the core and crust.  Both neutrons and
protons are superfluid in the core of this 1.5 $M_\odot$ neutron star.
In addition to the trough in the crust layers, the core develops a
complex structure.  All the sources of neutrino emission and the nucleon
heat capacity in the core are affected by the superfluidity, while the
electron heat capacity is not and becomes dominant in highly superfluid
cores.  As soon as the temperature drops significantly below the
critical temperatures $T_{cn}$ or $T_{cp}$, Cooper-pairing neutrino
emission comes into play.  It starts in the inner part of the core and
drives the temperature down.  At $t \sim 10^{-2}$ yr that region is even
cooler than in the non-superfluid model, while the other parts of the
core are slightly hotter.  By the age of 1 yr, this cool region includes
all of the core except the inner kernel.  As a result, thermal
relaxation proceeds on a shorter timescale, and at $t \la 20$ yr the
stellar interior is isothermal.  Thus, for the large assumed values of
$T_{cn}$ and $T_{cp}$, the neutrino emission due to Cooper pairing
becomes so strong that, instead of slowing down, the presence of the
core superfluidity accelerates the cooling.

Figure \ref{fig:c5_crsf} demonstrates the aforementioned effects on the
thermal relaxation time in the 1.5 $M_\odot$ neutron star.  The stellar
core is assumed to be non-superfluid.  The thermal relaxation stage
occurs about 2.6 times faster in the case of weak crust superfluidity
and about 3.5 times faster in the case of strong crust superfluidity,
compared to the non-superfluid crust.  And also, while the inclusion of
the neutrino emission due to Cooper pairing leads to faster cooling for
$t < 20$ yr, most of the accelerating effect is due to the reduction of
the crust heat capacity.  Therefore, the {\it relaxation time is greatly
reduced by the effect of superfluidity on heat capacity of free neutrons
in the stellar crust}.

\begin{table}
\caption{Normalized relaxation times 
         in the crust $t_1$, eq.\ (\protect\ref{cool-tw}),
         and in the core $t_2$, eq.\ (\protect\ref{eq:tcore})}
\begin{center}
\begin{tabular}{lrr}
\hline
                                     &  fast cooling    &  slow cooling \\
\hline
crust $t_1$ (yr) \\
\quad no SF                          &  $28.4 \pm 0.2$  &  $32.9 \pm 1.2$ \\
\quad weak crust SF                  &  $10.3 \pm 0.6$  &  $ 3.4 \pm 0.3$ \\
\quad weak core+crust SF             &  $11.5 \pm 0.5$  &  $25.0 \pm 0.4$ \\
\quad strong crust SF                &  $ 7.0 \pm 0.9$  &  $ 6.8 \pm 1.0$ \\
\quad strong core+crust SF           &  $ 6.2 \pm 0.4$  &  $ 5.7 \pm 0.2$ \\
core $t_2$ (yr) \\
\quad no SF                          &  $ 9.1 \pm 0.8$ \\
\quad weak core SF                   &  $11.2 \pm 0.5$ \\
\quad strong core SF                 &  $ 3.6 \pm 0.2$ \\
\hline
\end{tabular}
\label{tab:fits}
\end{center}
\end{table}

Still, the relaxation time of the superfluid crust satisfies the same
scaling relation, equation (\ref{cool-tw}), as for the non-superfluid
crust.  Table \ref{tab:fits} summarizes the results, taken as the mean
of $M = 1.5, 1.6, 1.7\, M_{\sun}$ models for fast cooling, and $M = 1.1,
1.2, 1.3, 1.4\, M_{\sun}$ models for slow cooling.  In case of the
fast-cooling models, crust superfluidity reduces the relaxation time by
up to a factor of four; $t_1 = 7.0 \pm 0.9$ yr for the strong
superfluidity.  In case of the slow-cooling models with strong crust
superfluidity the effect is similar, but with the weak superfluidity it
is much different, almost a factor of 10.  The inclusion of
superfluidity in the core does not affect the relaxation time
significantly.  The values of $t_1$ change by about $\approx 1$ yr and
are within the error-bars of the results for the non-superfluid cores.
The notable exception again is the weak superfluidity in the
slow-cooling models.

Superfluidity in the core changes the core relaxation time, although it
still scales as equation (\ref{eq:tcore}).  Weak core superfluidity only
increases the core heat capacity and delays the relaxation by a few
years, $t_2 = 11.2 \pm 0.5$ yr, while the strong superfluidity
suppresses the heat capacity and gives much faster relaxation, by a
factor 2.5.

The large difference of the values of $t_1$ for the slowly and rapidly
cooling stars with weakly-superfluid crusts indicates that, generally,
superfluidity may violate the simple scaling relations for the
relaxation time as well as the strict thermal decoupling of the crust
and the core prior to the relaxation.  This happens because
superfluidity makes the heat capacity (and other properties of matter) a
strong function of density.

The shortening of the thermal relaxation phase in a rapidly cooling
star, due to the superfluid reduction of the crustal heat capacity, was
emphasized by Lattimer et al.\ (1994).  Using a model of superfluidity,
which is closer to our model of strong superfluidity in the crust,
Lattimer et al.\ find that the relaxation time becomes three times as
short, $t_w = 8.4 \pm 2.0$ yr, in qualitative agreement with our
results.  Our calculations indicate that the effect is sensitive to the
model of neutron superfluidity in the crust and thus, it can be used to
test such models.

When thermal relaxation is over, the cooling is mainly regulated by the
neutrino luminosity and heat capacity of the core.  The processes in the
crust cease to play a significant role except for the very low-mass
neutron stars with large crusts.  We find, however, that for some
superfluid models the neutrino luminosity of the crust may affect the
cooling for a short period of time during the transition from the
neutrino cooling era to the photon era.  For instance, cooling of
superfluid neutron stars after the thermal relaxation is discussed in
detail by Yakovlev et al.\ (1999).

\section{Conclusions}
  \label{conclude}

We have studied thermal evolution of young neutron stars using a new
cooling code and updated physics input.  The effective surface
temperature of isolated neutron stars in the first 100 yr is determined
by the properties of the crust.  After the cooling wave from the core
reaches the surface, the effective temperature drops from 250 eV to 50
eV or lower.  We confirm the conclusion of Lattimer et al.\ (1994) that
the duration of the relaxation epoch in a fast cooling scenario depends
sensitively on the heat capacity and thermal conductivity of the crust.
The relaxation time scales with the size of the crust,
eq. (\ref{cool-tw}), and the normalization is $t_1 = 28$ yr without
superfluid effects or $t_1 = 7$ yr with strong superfluid effects.  We
find that the same conclusion holds in the case of slow cooling, in
which thermal relaxation is less pronounced.

We also investigate the effects of various neutrino emission mechanisms
in the crust, superfluidity of free neutrons, heat capacity, and thermal
conductivity on the relaxation time (as summarized in Table
\ref{tab:lab} and Fig. \ref{fig:twvar}).  The relaxation time is most
sensitive to the variations of $\kappa$ and $C_v$ in the density range
$10^{13} - 1.5\times 10^{14}$ g cm$^{-3}$ near the crust-core interface.

Young cooling neutron stars can serve as astrophysical laboratories of
matter at subnuclear densities.  The sensitive dependence of the
relaxation time on the microscopic properties of matter in the deep
inner crust provides a possibility to study these properties by
observing the emergence of the cooling wave.  In order to realize this
method, one needs to detect thermal emission from a very young neutron
star in the range from 50 to 250 eV.  Such stars have not been found so
far, but may be detected in young supernova remnants by Chandra and
XMM.

To improve the cooling theory of young neutron stars, it would be
desirable to continue theoretical studies of microscopic properties of
matter in the inner crust, in particular, the possible unusual phases of
nonspherical nuclei at $\rho > 10^{14}$ g cm$^{-3}$, as well as the
thermal conductivity, heat capacity, and neutrino emission of such
matter.  It is also important to refine our knowledge of matter
containing spherical nuclei at $\rho \la 10^{14}$ g cm$^{-3}$.  In this
paper we present (Appendix) and use the results of new calculations of
the electron thermal conductivity due to the scattering of electrons off
spherical atomic nuclei, taking into account the effects of finite sizes
of proton nucleus cores.

\bigskip\noindent
{\bf Acknowledgements}
\\
We are grateful to P.\ Haensel and A. D.\ Kaminker for stimulating
discussions and useful suggestions, and to D.\ Page for his suggestion
to optimize the performance of the cooling code.  This work was
supported in part by RFBR (grant No.\ 99-02-18099), INTAS (grant No.\
96-0542), NSF (grant No.\ PHY99-07949), and PPARC.

\newcommand{\req}[1]{equation (\ref{#1})}
\newcommand{\gcc}{{\rm~g\,cm}^{-3}}
\newcommand{\beq}{\begin{equation}}
\newcommand{\eeq}{\end{equation}}
\newcommand{\bea}{\begin{eqnarray}}
\newcommand{\eea}{\end{eqnarray}}
\newcommand{\Compton}{\lambda\hspace{-.54em}{\mbox{$^-$}}}

\appendix
\section{Electron conductivity of the inner crust}

In the inner crust, the electrons are strongly degenerate.  The heat and
charge are transported mainly by the electrons with energies in a narrow
thermal band near the Fermi level $\epsilon_{\rm F}$.  The coefficients
of the electrical and thermal conductivities can be written as
\beq
  \sigma  =  {e^2 n_{\rm e} \tau_\sigma \over m_{\rm e}^\ast},
\quad
   \kappa  =  { \pi^2 k_{\rm B}^2 T n_{\rm e} \tau_\kappa
                 \over 3 m_{\rm e}^\ast},
\label{dg-kin-coeff}
\eeq 
where $m_{\rm e}^\ast \equiv \epsilon_{\rm F}/c^2$, $n_{\rm e}$ is the
electron number density, $k_{\rm B}$ is the Boltzmann constant, and
$\tau_\sigma$ and $\tau_\kappa$ are the effective relaxation times.  In
the presence of several electron scattering mechanisms, the cumulative
relaxation time is determined by the Matthiessen rule (e.g., Ziman
1960).  In the envelopes of neutron stars, one traditionally considers
the electron-ion ($ei$), electron-electron, and electron-impurity
scatterings.  We focus on the main mechanism, the $ei$ scattering.

A modern theoretical treatment of the conductivities due to the $ei$
scattering in strongly coupled Coulomb plasmas of ions (atomic nuclei)
has been proposed by Baiko et al.\ (1998) and applied to the outer
envelopes of neutron stars by Potekhin et al.\ (1999).  In the latter
work, practical formulae for $\sigma$ and $\kappa$ incorporating all
three scattering mechanisms have been derived; the corresponding Fortran
code is available electronically at
http://www.ioffe.rssi.ru/astro/conduct/.

For the inner crust these expressions must be modified.  First, the
atomic nuclei cannot be considered as point-like.  Second, the $ei$
scattering in the inner crust at temperatures much below $10^8$ K
changes its character: so called {\em umklapp\/} processes cease to
dominate and the {\em normal\/} processes (with electron momentum
transfer within one Brillouin zone) become more important.

\subsection{Effects of finite size of ions}
  \label{sec-A1}

The effective $ei$ relaxation time is
\beq
   \tau_{ei}= { 3 \pi \hbar \over 
   4 Z \epsilon_{\rm F} \alpha_{\rm f}^2 \Lambda_{ei}(\epsilon_{\rm F}) },
\label{ei-tau}
\eeq
where $\alpha_{\rm f} \equiv e^2/(\hbar c)$ is the fine structure
constant, and $\Lambda_{ei}$ is the Coulomb logarithm.  For a classical
plasma of ions in the Born approximation (e.g., Potekhin et al.\ 1999),
\bea
    \Lambda_{ei} & = & \int_0^{2k_{\rm F}}
    {\rm d}q\; q^3\, |\phi(q)|^2 \, S(q) \,
    \left[1 - \beta_{\rm r}^2
    \left({q\over 2 k_{\rm F}}\right)^2  \right], 
\label{ei-L}
\\&&\qquad
    \phi(q) \equiv { F(q)  \over q^2 \, \varepsilon(q)},
\nonumber
\eea
where $k_{\rm F} = (3 \pi^2 n_e)^{1/3}$ is the electron Fermi wave
number, $\beta_{\rm r} = v_{\rm F}/c$, $v_{\rm F} = \hbar k_{\rm
F}/m_{\rm e}^\ast$ is the Fermi velocity, $\varepsilon(q)$ is the static
dielectric function, $S(q)$ is the static structure factor of the ions
(more precisely, its inelastic part -- see Baiko et al.\ 1998), and
$F(q)$ is the form factor of the ions.  The latter three functions allow
for the electron polarization, ion-ion correlations, and the finite size
of the ions, respectively.

The quantization of ionic motion becomes important at $T \ll T_{\rm p}$,
where $T_{\rm p} = \hbar\,\sqrt{4 \pi n_{\rm i} Z^2 e^2/ M} \,/k_{\rm
B}$ is the ion plasma temperature, $n_{\rm i} = n_{\rm e}/Z$ is the ion
number density, $M = A \, m_{\rm u}$ is the ion mass, and $m_{\rm u}$ is
the atomic mass unit.  In this case, in \req{ei-L} one should use an
{\em effective\/} structure factor, different for $\sigma$ and $\kappa$.
These effective factors have been derived by Baiko et al.\ (1998) and
fitted by Potekhin et al.\ (1999) for the case where the umklapp
processes dominate.

\subsubsection{Uniform charge distribution in atomic nuclei}

At $\rho \ll \rho_{cc}$ a good approximation for the charge density
within a nucleus is a step function.  The corresponding form factor is
\begin{equation}
  F(q) = \frac{3}{(q r_{\rm nuc})^3}
   \left[ \sin(q r_{\rm nuc}) - q r_{\rm nuc} \cos(q r_{\rm nuc}) \right],
\label{formf}
\end{equation}
where $r_{\rm nuc}$ is the proton core radius.

In general, $\Lambda_{ei}$ depends on
$\rho$, $T$, $Z$, $A$, $r_{\rm nuc}$,
and it appears to be almost independent of the type of Coulomb lattice.
It is convenient, however, to introduce
the dimensionless parameters: 
the electron relativity parameter $x_{\rm r} = \hbar k_{\rm F}/(m_{\rm e}c)$;
the ion size parameter $x_{\rm nuc} = r_{\rm nuc}/a_{\rm i}$,
where $a_{\rm i} = (4 \pi n_{\rm i}/3)^{-1/3}$ is the ion-sphere radius;
the Coulomb coupling parameter of ions 
$\Gamma = Z^2 e^2 /(a_{\rm i} k_{\rm B} T)$;
the inverse ion quantum parameter 
$t_{\rm p}\equiv T/T_{\rm p}$;
and the electron screening parameter
$s_{\rm e}=k_{\rm TF}^2/(2k_{\rm F})^2=\alpha_{\rm f}/(\pi\,\beta_{\rm r})$,
where $k_{\rm TF}$ is the electron screening wave number.
In accordance with the previous results,
it is also convenient to introduce an auxiliary parameter
$s_{\rm D}=(2k_{\rm F}\,r_{\rm D})^{-2}$,
where $r_{\rm D}=a_{\rm i}/\sqrt{3\Gamma}$ is the Debye screening length 
for the ideal plasma of ions.
Finally, the basic parameters that characterize
the Coulomb lattice are the normalized first- and second-negative moments
of the phonon frequency, $u_{-1}\approx3$ and $u_{-2}\approx13$.

We have calculated numerically the effective Coulomb logarithms
$\Lambda_{ei}^{\sigma,\kappa}$ for about a hundred pairs ($\rho,T$),
varying over orders of magnitude in the domain of strongly degenerate
electrons and strongly coupled ions, for eight ion species from $Z=12$
to $Z=60$, and for five values of $x_{\rm nuc}$ from 0 to 0.4.  We have
included the non-Born corrections 
in the same manner as Yakovlev (1987).  The results can be
fitted using the {\em effective screening function}
\beq
       \left| \phi^{\rm eff}(q) \right|^2 =
   { 1 \over (q^2+q_s^2)^2}
   \, [1-{\rm e}^{-s_0 q^2} ] \, 
  \, {\rm e}^{- s_1 q^2} \,
       G_{\sigma,\kappa} \, D
\label{Ueff}
\eeq
instead of $|\phi(q)|^2 S(q)$ in \req{ei-L}.  Here, the first term
corresponds to the Debye screening with the effective screening wave
number $q_s$, the factor in square brackets corrects for the ion
correlations, and the functions $G$ and $D$ describe ionic quantum
effects, as in Potekhin et al.\ (1999).  An additional factor ${\rm
e}^{- s_1 q^2}$ plays a role of the effective form factor.  The
numerical values of $\Lambda$ obtained from the accurate theory are
reproduced (within several percent in the most important $\rho$--$T$
range) if we use the effective screening function (\ref{Ueff}) with the
following parameters:
\begin{eqnarray}
&&\hspace*{-1.5em}
     s \equiv q_{\rm s}^2 /(2k_{\rm F})^2 =
    (s_{\rm i} + s_{\rm e})\,{\rm e}^{-\beta_Z},
\\&&\qquad
\beta_Z=\pi\alpha_{\rm f} Z\beta_{\rm r},
\qquad
     s_{\rm i} = s_{\rm D}\,(1+0.06\,\Gamma) \, {\rm e}^{-\sqrt\Gamma};
\nonumber
\\ &&\hspace*{-1.5em}
    w \equiv (2k_{\rm F})^2\,s_0 = (u_{-2}/s_{\rm D}) \,(1+\beta_Z/3);
\\ &&\hspace*{-1.5em}
    w_1 \equiv (2k_{\rm F})^2\,s_0 
\nonumber\\&&
= 14.73 \, x_{\rm nuc}^2\, 
        \left(1 + \sqrt{x_{\rm nuc}}\,\,Z/13 \right)\,(1+\beta_Z/3),
\label{s1}
\\ &&\hspace*{-1.5em}
    G_\sigma = 
       (1 + 0.0361\, Z^{-1/3}/t_{\rm p}^2)^{-1/2} \,(1+0.122\,\beta_Z^2), 
\\ &&\hspace*{-1.5em}
    G_\kappa = G_\sigma +
      { 0.0105\, t_{\rm p} \over (t_{\rm p}^2 + 0.0081)^{3/2}}
        \left[ 1+\beta_{\rm r}^3 \beta_Z \right]
      \left( 1-Z^{-1} \right) 
\nonumber\\&&\qquad\times
         ( 1+x_{\rm nuc}^2\,\sqrt{2Z} ) ,
\label{Gkappa}
\\ &&\hspace*{-1.5em}
D = \exp\left[ - 0.42\,\sqrt{x_{\rm r}/ (AZ)}\, u_{-1} \, 
                         \exp(-9.1 t_{\rm p})\right].
\end{eqnarray} 
If $x_{\rm nuc}=0$, these equations reduce to those
presented in Potekhin et al.\ (1999).

With the effective screening function in the form (\ref{Ueff}),
we can integrate \req{ei-L} analytically and obtain
\beq 
        \Lambda_{ei}^{\sigma,\kappa} = 
               \left[ \Lambda_0(s,w+w_1) - \Lambda_0(s,w_1) \right] \,
       G_{\sigma,\kappa} \, D,
\label{Lambdafit} 
\eeq 
where 
\begin{eqnarray} 
&&\hspace*{-1.5em}
    \Lambda_0(s,w) = \Lambda_1(s,w) - 
        \beta_{\rm r}^2 \, \Lambda_2(s,w),
\\&&\hspace*{-1.5em}
       2\, \Lambda_1(s,w) = \ln{s+1 \over s} + {s\over s+1} 
          \, (1-e^{-w})  
\nonumber\\ &&\qquad
     - (1+s w) \, e^{s w}  
     \left[ {\rm E}_1\,(s w) - {\rm E}_1\,(sw + w) \right], 
\label{Lambdafit1}
\\ &&\hspace*{-1.5em}
  2\, \Lambda_2(s,w) = {e^{-w} - 1 + w \over 
        w} - {s^2 \over s+1} \, 
           (1-e^{-w}) - 2 s \ln{s+1 \over s} 
\nonumber\\ && \qquad
     +   s  \,(2 + s w) \, e^{s w}  
    \left[ {\rm E}_1\,(s w) - {\rm E}_1\,(s w + w) \right], 
\label{Lambdafit2}
\end{eqnarray} 
and ${\rm E}_1(x) = \int^{\infty}_x y^{-1}e^{-y}{\rm\,d}y$ is the
standard exponential integral.  Note that using equations
(\ref{Lambdafit1}) and (\ref{Lambdafit2}) directly 
may result in large numerical round-off errors
in the limiting cases $s \ll 1$, $w \ll 1$, or $w \gg 1$. In these cases,
explicit asymptotic expressions of Potekhin et al. (1999) should be
used.

\subsubsection{Realistic charge distribution}

Near the bottom of the crust, the boundary of the proton core in nuclei
becomes fuzzy and the above results should be modified.  Oyamatsu (1993)
approximated the proton charge distribution by a function proportional
to $(1-r/r_{\rm m})^b$ (at $r<r_{\rm m}$), where the power index $b$
controls the ``sharpness" of the charge profile.  The parameters $b$ and
$r_{\rm m}$ have been described by the simple functions of mass density
in the smooth composition model of crust matter (Kaminker et al.\ 1999).
Using that model we have calculated the electrical and thermal
conductivities for $\rho = 10^9 - 10^{14.1} \gcc$ and $T = 10^7 - 10^9$
K, and compared the results with the fitting formulae
(\ref{Lambdafit})--(\ref{Lambdafit2}).  We have found that the
electrical conductivity is reproduced (within $\approx5$--10\%), if we
define $x_{\rm nuc}$ in \req{s1} so as to reproduce the same {\em
mean-square\/} proton-core radius as in the approximation of uniform
charge distribution:
\beq
   x_{\rm nuc} = {r_{\rm m}\over a_{\rm i}}\,
{1-15/(5+b)+15/(5+2b)-5/(5+3b)\over 1-9/(3+b)+9/(3+2b)-1/(1+b)}.
\eeq
In \req{Gkappa} which determines the thermal conductivity at $T\ll
T_{\rm p}$, one should additionally multiply $x_{\rm nuc}$ by a factor
$b/(0.5+b)$. The maximum fit error of $\kappa$ does not exceed 13\%.

\subsection{Normal processes at very low temperatures}
  \label{sect-umklapp}

Near the boundaries of the Brillouin zones the dispersion relation of
electrons differs from the free-electron case, and at the boundaries the
electron energy spectrum contains gaps.  The gaps $\Delta\epsilon$ can
be estimated in the weak coupling approximation (e.g., Kittel 1986) as
$\Delta\epsilon \sim \phi(k_{\rm F})=4\pi Z e^2 n_{\rm i} k_{\rm
F}^{-2}=4e^2/(3\pi k_{\rm F})$.  The effect of gaps is most significant
if the deviation of the electron momentum from the intersection line
between the Fermi surface and the Brillouin zone boundary does not
exceed $\Delta k \sim \Delta\epsilon/(\hbar v_{\rm F})
\sim \frac{4}{3\pi}\,(\alpha_{\rm f}/\beta_{\rm r})\,k_{\rm F}\ll k_{\rm F}$.
However, with decreasing temperature the strips of the Fermi surface,
between which the umklapps proceed effectively, become narrower and
closer to these intersection lines.  When the widths of the strips,
$\sim t_{\rm p} \,(6\pi^2 n_{\rm i})^{1/3}$, become smaller than $\Delta
k$, the umklapp processes are frozen out and the normal processes
prevail.  The above estimates indicate that this happens when the
temperature falls below
\beq
  T_{\rm u} \sim T_{\rm p} Z^{1/3} \alpha_{\rm f}/(3\beta_{\rm r}).
  \label{Tu}
\eeq

In this case, the formalism used in Sec. \ref{sec-A1} becomes invalid.
The asymptotes for the $ei$ scattering rates at $T \ll T_{\rm u}$ have
been obtained by Raikh \& Yakovlev (1982). In our notation, they yield:
\beq
\left\{\begin{array}{c}
       \Lambda_{ei}^\sigma \\ \Lambda_{ei}^\kappa
\end{array} \right\}  =
{a_\zeta\,x_{\rm r}^{1/2} \over A^{1/2}\,Z}\times
\left\{\begin{array}{c}
       (4/3)\,(\alpha_{\rm f}/\beta_{\rm r})\,t_{\rm p}^5 \\ t_{\rm p}^3
\end{array} \right\} ,
\label{ei-L-low}
\eeq
where 
\[
     a_\zeta = 180\,\left({3\over\pi\alpha_{\rm f}}\,{m_{\rm e}\over m_{\rm u}}\
         \right)^{1/2} \zeta(5)   \approx     50,
\]
and $\zeta(5)=1.0369$ is the value of the Riemann zeta function.

Now we interpolate between the
high-temperature Coulomb logarithm $\Lambda_{ei,{\rm
high}}^{\sigma,\kappa}$ given by \req{Lambdafit} and the low-temperature
asymptote $\Lambda_{ei,{\rm low}}^{\sigma,\kappa}$ given by
\req{ei-L-low}:
\bea
    \Lambda_{ei}^{\sigma,\kappa} &=& 
        \Lambda_{ei,{\rm high}}^{\sigma,\kappa} \exp(-T_{\rm u}/T)
\nonumber\\&&
 + \Lambda_{ei,{\rm low}}^{\sigma,\kappa} \,
\left[ 1 - \exp(-T_{\rm u}/T) \right].
\label{ei-L-complete}
\eea

\begin{figure}
\begin{center}
\leavevmode
\epsfxsize=8cm 
\epsfbox{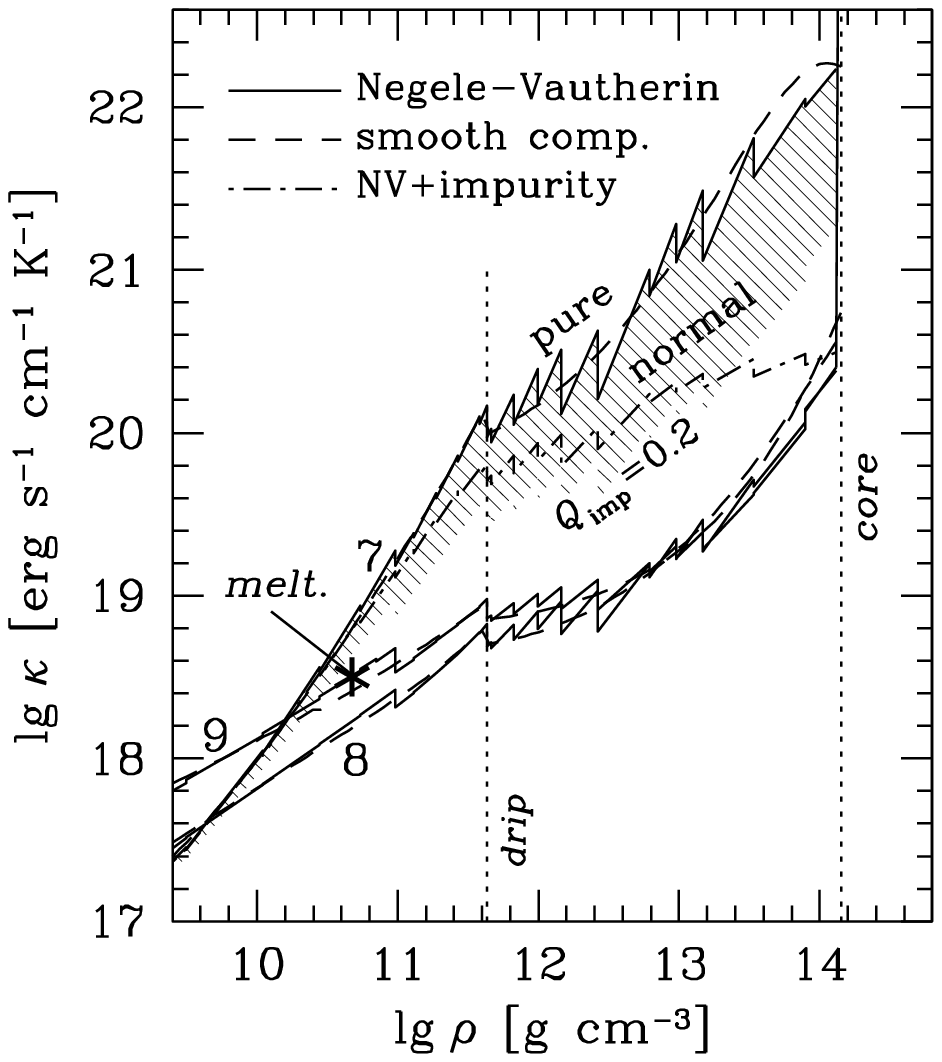} 
\end{center}
\caption[ ]{
Density dependence of the electron thermal conductivity at $T=10^7$,
$10^8$ and $10^9$ K in the neutron star crust.  Solid and dot-dashed
lines: ground state composition (Negele \& Vautherin 1973), rectangular
profile of nuclear charge (solid lines: pure crystal or liquid;
dot-dashed line: including impurities; 
bottom of the hatched region: 
neglecting freezing of umklapp processes
at $T=10^7$ K); dashed
lines: smooth-composition model, realistic profile of nuclear charge, no
impurities.
}
\label{fig-condapp}
\end{figure}

Figure \ref{fig-condapp} shows the thermal conductivities for three
values of the temperature.  In the neutron star envelope, the plasma is
in the solid phase except for the segment to the left of the asterisk at
the highest $T=10^9$ K.  At the lowest $T=10^7$ K, the change in the
scattering from umklapp to normal processes leads to a significant
increase in the conductivity shown by the hatched region.  However, this
increase can be compensated by the scattering off impurities.  For
example, the dashed line shows the conductivity for $Q_{\rm imp} \equiv
\langle (Z_{\rm imp}-Z)^2\,n_{\rm imp}\rangle /n_{\rm i} = 0.2$, where
$n_{\rm imp}$ and $Z_{\rm imp}$ are the impurity number density and
charge number, respectively.  For comparison, dashed lines show the
thermal conductivity in the smooth-composition model with smooth proton
charge profile.

\end{document}